\documentclass[5p]{elsarticle}

\usepackage{epsfig}
\usepackage{dcolumn}
\usepackage{bm}
\usepackage{subfig}
\usepackage{float}
\usepackage{amssymb}
\usepackage{natbib}

\newcommand{\eq}[1]{
\vspace{-0.3cm}
\begin{eqnarray}
	#1
\end{eqnarray}
}
\newcommand{\Sch}{Schr\"odinger }

\begin{document}

\bibliographystyle{model1-num-names}

\title{Probing New Limits for the Violation of the Equivalence Principle in the solar-reactor neutrino sector as a next to leading order effect}

\author{G. A. Valdiviesso}
\ead{gustavo.valdiviesso@unifal-mg.edu.br}
\address{Instituto de Ci\^encia e Tecnologia\\ Universidade Federal de Alfenas, Unifal-MG \\ Av. Jos\'e Aur\'elio Vilela, 11999 \\ 37715-400 Po\c cos de Caldas MG Brazil}

\author{M. M. Guzzo}
\ead{guzzo@ifi.unicamp.br}

\author{P. C. Holanda}
\ead{holanda@ifi.unicamp.br}
\address{Instituto de F\'\i sica Gleb Wataghin\\ Universidade Estadual de Campinas, UNICAMP \\ Rua S\'ergio Buarque de Holanda, 777 \\ 13083-859 Campinas SP Brazil}

\begin{abstract}
New limits for the Violation of Equivalence Principle (VEP) are obtained considering the mass-flavor mixing hypothesis. 
This analysis includes observations of solar and reactor neutrinos and has obtained a limit for the VEP parameter $|\Delta \gamma|$ 
contributing to the $\nu_e$ and $\bar{\nu}_e$ disappearance channels of the order $|\Delta \gamma|<10^{-14}$, when it is assumed that 
neutrinos are mainly affected by the gravitational potential $\phi\approx 10^{-5}$ due to the Great Attractor. 

\end{abstract}


\maketitle

\section{\label{sec:introduction}Introduction}

Two decades ago, Gasperini introduced the idea of mixing between flavor and gravitational
neutrino eigenstates,  leading to a Violation of the Equivalence Principle
(VEP)~\citep{Gasperini}. The purpose of such model was to find a solution to  the solar
neutrino problem through an oscillation mechanism ``\`a la" Pontecorvo induced by a minimal
coupling between the  gravitational field $\phi(x)$ and the neutrino field.  

This approach just considered the neutrino kinetic energy content as its mass when coupling
with gravity. Later, Halprin and  Leung~\citep{Halprin1,Halprin2} introduced independently a
neutrino field coupled with a linearized space-time  metric, such that $g_{\mu\nu}\equiv
\eta_{\mu\nu}+h_{\mu\nu}$ where $\eta_{\mu\nu}=diag(1,-1,-1,-1)$ and~\citep{Will}
$h_{\mu\nu}=-2\phi(x)\delta_{\mu\nu}$.

Although these hypotheses have been formulated for massive neutrinos, no experimental data
available at that time could distinguish  between mass-flavor and gravity-flavor oscillations.
It was much simpler therefore to consider only one of these two different effects.  As a
consequence, experimental confrontation made before the first KamLAND results~\citep{KamLAND}
considered this simple case of massless  neutrinos, mixed only via gravitational interaction
(see however~\citep{Iida:1992vh,Minakata:1994kt,Minakata:1996nd} for a treatment with  mass and
VEP effects). In fact, a ``just so" vacuum solution could explain all solar data. On the other
hand, the increasing evidence of  neutrino disappearance at short distances ($\cong 180km$)
cannot be described by this kind of solution, which leads to neutrino oscillation  lengths of
the order of the Sun - Earth distance.

With the increasing statistics on neutrino coming from the Sun, reactors and the
accumulated data from all other sources, one could ask what limits can now be imposed to VEP
parameters when we assume the mass-flavor mixing and MSW mechanism added by gravitational VEP
interactions in a neutrino system. In other words, would neutrinos be good probes for
effects coming from VEP?

The VEP phenomenon manifests itself as a difference in the gravitational coupling for different
states. In order to parametrize its effects we will adopt the Post-Newtonian
Parametrization~\citep{Gravitation}, where any difference from known gravitational Newtonian
constant $G_N$ is included in a $\gamma$ factor, so that $G_N' = \gamma_m\ G_N$, where
$\gamma_m\equiv\gamma(m)$ depends on the mass $m$ of the system.  Once that the $G_N$ constant
is already been considered in the definition of the gravitational potential $\phi(r)$, one may
also define the $\gamma$ factor as:

\eq{\label{eq:gammaphi} \phi' = \gamma_m\ \phi\ , }

\noindent where $\phi$ is defined to be positive. For macroscopic bodies $A$ and $B$, the
difference between their $\gamma_A$ and $\gamma_B$ factors $\Delta \gamma = \gamma_A -
\gamma_B$ has been measured with free fall experiments. Several gravitational sources are
considered: the Sun, the Earth,  and the galactic center, obtaining a superior limit
$\Delta\gamma<10^{-12}$ for astrophysical sources~\citep{Su} and $\Delta\gamma<10^{-9}$ for
terrestrial  experiments~\citep{Smith}. Interesting enough, some astrophysical events like
pulsars with peculiar frequencies, could be explained if the neutrinos were experiencing 
VEP~\citep{Horvat, Barkovich}. On the other hand, neutrinos cannot violate the equivalence
principle by more than $1$ part in $100$ ($90\%C.L.$)~\citep{Damour3}  since one would be
observing more erratic pulsars. Limits on VEP have also been obtained in neutrino oscillations 
experiments\citep{Concha}, although for a different set of parameters than the ones we are 
analyzing in this Letter. A comparison of all these limits will be done in section \ref{conclusao}.

\section{VEP Model for Massive Neutrinos}

We start stating that the model will apply only to weak gravitational fields, 
so that no spin-gravity effects will be considered here. 
By doing so, one may use the Klein-Gordon equation to describe the neutrino field: 

\eq{\left(g_{\mu\nu}\partial^{\mu}\partial^{\nu}+m^2\right)\Psi =0,} 

\noindent where $g_{\mu\nu}$ is the metric tensor and $\Psi$ represents the neutrino field.

Following Halprin's approach, the metric tensor for a weak field can be written 
as $g_{\mu\nu}=\eta_{\mu\nu}+h_{\mu\nu}(x)$ where  $\eta_{\mu\nu}=diag(1,-1,-1,-1)$ 
and $h_{\mu\nu}=-2\gamma_m\phi(x)\delta_{\mu\nu}$~\citep{Will}, where the redefinition (\ref{eq:gammaphi}) 
is being used from now on. So, the Klein-Gordon equation with weak gravitational 
field is $\left[\left(\eta_{\mu\nu}-2\gamma_m\phi(x)\delta_{\mu\nu}\right)\partial^{\mu}\partial^{\nu}+m^2\right]\Psi =0$. 
Assuming a plane-wave solution of the form $\Psi=\Psi_0 e^{i(\vec{p}\cdot\vec{x}-Et)}$, 
one arrives at the energy-momentum relation for this interacting system: 
$E^2(1-2\gamma_m\phi)= p^2(1+2\gamma_m\phi) + m^2$. Using the fact that for neutrinos $m\ll p$ 
and ignoring terms with order higher than $O(\phi^2)$, we finally have the energy-momentum 
relation for small masses and weak gravitational potential:

\eq{
E &\cong &  p\ (1+2\gamma_m\phi) +\frac{m^2}{2p}(1+4\gamma_m\phi)\ \ . \label{E_Halprin}
}

\noindent The above expression can be re-written as $E=E_m+E_g$ so that $E_{m}=p+\frac{m^2}{2p}$ is the free-particle energy-momentum 
relation (with $m\ll p$) and $E_{g}= 2\gamma_m\phi\left(p+\frac{m^2}{p}\right)$ is the gravitational contribution to the total energy. 

A remark is in order: to introduce the neutrino mixing, one has to define a basis on which each phenomenon takes place. The most general scheme for this model 
would be a three basis system: a physical basis (states with definite mass), a weak basis (states with definite flavor) and a gravitational basis 
(states with definite gravitational couplings). This would mean that the dynamical and gravitational contributions to the total energy, $E_{m}$ and $E_{g}$, 
could not be simply added any more. Instead, the two physical quantities should be assigned to operators on different bases. 
Considering the further inclusion of weak interactions, and one third basis for it, the model will end with five free parameters~\citep{Halprin2} 
(considering only two neutrino flavors). Although it is possible to carry on such analysis, it is interesting to test simpler models and, 
if any signal of VEP is found, a more complete analysis could be made in future works. To obtain a simpler model, we follow the hypothesis that the 
gravitational interaction takes place on the physical mass basis. This is exactly what has been done until now, when deriving the expression (\ref{E_Halprin}). 

Considering only two neutrino flavors, each mass eigenstate has total energies $E_1$ and $E_2$, given by expression~(\ref{E_Halprin}), 
using $m\rightarrow m_1$ and $\gamma_m\rightarrow\gamma_1$ for $E_1$, so that:

\eq{
E_1 &= &  p\ (1+2\gamma_1\phi) +\frac{m_1^2}{2p}(1+4\gamma_1\phi)\ ~~,
\label{E_1}
}

\noindent and $m\rightarrow m_2$ and $\gamma_m\rightarrow\gamma_2$ for $E_2$: 
\eq{
E_2 &= &  p\ (1+2\gamma_2\phi) +\frac{m_2^2}{2p}(1+4\gamma_2\phi)\ \ . \label{E_2}
}

\noindent To describe a two level system, we introduce the Hamiltonian

\eq{\label{H_massa2}
H^{(m)}=\frac{\Delta E}{2}\left(
\begin{array}{cc}
-1 & 0 \\
0 & 1
\end{array}
\right)\ 
}

\noindent where $\Delta E= E_2 - E_1$ such that

\eq{
\Delta E & =&  \frac{\Delta m^2}{2p} + 2p\ \phi\Delta \gamma + \frac{\phi}{p}\left(\bar{\gamma}\Delta m^2 + \bar{m}^2\Delta \gamma\right)\label{DeltaE-completo}
}

\noindent where $\Delta m^2=m_2^2-m_1^2$, $\Delta\gamma=\gamma_2-\gamma_1$, $\bar{\gamma}=(\gamma_2+\gamma_1)/2$ and
$\bar{m}^2=(m_2^2+m_1^2)/2$.

Not all of these terms will contribute. Of the three terms with dependence on $1/p$ in (\ref{DeltaE-completo}), 
the last two ones are negligible, mainly because of the potential $\phi$. Comparing all the sources of gravity 
that might have some effect here, as the Earth, the Sun, and larger scale structures such the Great Attractor ~\citep{Burstein,Kraan}, 
the last contributes most, imposing a practically constant potential $\phi\approx 3\times 10^{-5}$ ~\citep{Kenyon}, that is at least
one order of magnitude larger than the other ones~\citep{Halprin2}. Using the definition of $\gamma$ in (\ref{eq:gammaphi}) and other 
VEP tests already cited, then $\bar{\gamma}\cong 1$ with $\Delta \gamma<10^{-9}$. 
These statements assure that $\phi\left(\bar{\gamma}\Delta m^2 + \bar{m}^2\Delta \gamma\right)<<\Delta m^2$, so that $\Delta E$ may be considered only as

\eq{
\Delta E &\cong&  \frac{\Delta m^2}{2p} + 2p\ \phi\Delta \gamma ,\label{DeltaE} \\
&\equiv& \frac{\Delta_G}{2E}
}

\noindent where the usual consideration for neutrinos $p=E$ was used and $\Delta_G = \Delta m^2 + 4E^2\phi\Delta\gamma$ is defined as an effective mass scale. 

We assume $m_2>m_1$. Nevertheless, the same hierarchy does not have to hold for $\gamma_1$ and $\gamma_2$. 
Previous models for VEP considered only gravitational states for massless neutrinos. 
Consequently $\gamma$'s could arbitrarily follow the hierarchy $\gamma_2>\gamma_1$ in the same way it was done for the masses. 
In a model with VEP and massive neutrinos, the $\gamma$'s are dependent on the masses and so it is clear from the definition of $\Delta_G$ that the 
hierarchy between $\gamma_1$ and $\gamma_2$ will have influence on the resulting phenomenology.  So we must consider two possibilities: if $\gamma_2>\gamma_1$, 
following the same relationship defined for the masses, then $\Delta \gamma>0$; if $\gamma_2 < \gamma_1$, then an inversed hierarchy on the VEP sector appears, 
and then $\Delta \gamma<0$. We consider

\eq{\label{Delta_Gpm}
\Delta_{G} = \Delta m^2 \pm 4E^2 \left|\phi\Delta \gamma \right|\ ,
}

\noindent where $\left|\phi\Delta \gamma \right|$ is one single parameter of the model and no further discussions about the single value of $\phi$ are needed, 
as long as it is considered as a constant. The two possible hierarchies between the $\gamma$'s will be referred simply as $+VEP$ and $-VEP$ 
for the plus and minus sign on (\ref{Delta_Gpm}), respectively. 

Note that the particular case when $E=E_\ast$, where

\eq{\label{EG}
E_{\ast}=\frac{1}{2}\sqrt{\frac{\Delta m^2}{\left|\phi\Delta \gamma \right|}}\ ,
}

\noindent implies $\Delta E=0$ for $-VEP$ case and the mass eigenstates become degenerate. On the contrary, $\Delta E$ never vanishes for $+VEP$, 
but it presents a minimum value exactly for $E=E_\ast$ defined in eq. (\ref{EG}), i. e., 

\eq{\label{dEG}
\left.\frac{d}{dE}\Delta E\right|_{E=E_\ast}=0
}

\noindent for $+VEP$. Therefore $E=E_\ast$ is a critical energy of the model, either for $+VEP$ and $-VEP$ cases.

As the energy $E$ is a constant of motion, any previous solutions of the standard neutrino mixing model can accommodate the VEP hypothesis only by 
doing the substitution $\Delta m^2\rightarrow\Delta_G$.
Furthermore, no mention of the weak basis mixing was needed until this point. In this simplified two-bases version of the VEP model, 
gravity has no influence over the vacuum mixing (which would not be the case if a three-bases model is considered~\citep{Halprin2}).

The evolution of flavor states is given by the \Sch equation $i\frac{d}{dt}\Psi^{(f)}=H^{(f)}\ \Psi^{(f)}$, with 

\eq{
\Psi^{(f)} & = & U\ \Psi^{(m)}\ \mbox{ and}\label{eq:Psifm} \\
H^{(f)} &=& U^\dagger H^{(m)}\ U \label{eq:Hfm}
}

\noindent where $\Psi^{(f)}$ and $H^{(f)}$ represent the states and the Hamiltonian, written on the flavor basis $(f)$. In general, 
states and operators in both bases are related through expressions (\ref{eq:Psifm}) and (\ref{eq:Hfm}) respectively, where $U$ is a $SU(2)$ transformation. 
When dealing with only two bases, $U$ has only one physical parameter~\citep{smirnov,Halprin2} which can be expressed as

\eq{\label{eq:U}
U=
\left(
\begin{array}{cc}
\cos \theta & \sin \theta \\
-\sin \theta & \cos \theta
\end{array}
\right)\ .
}

In the flavor basis, one can introduce the effective weak potential~\citep{W}:

\eq{\label{eq:V-weak}
V_W^{(f)}=
\frac{\sqrt{2}}{2}G_F n_e\left(
\begin{array}{cc}
1 & 0 \\
0 & -1
\end{array}
\right)\ ,
}

\noindent that describes the influence of a material medium on the neutrino conversion, known as the MSW effect~\citep{MS}. 
In expression (\ref{eq:V-weak}), $G_F$ is the Fermi constant and $n_e\equiv n_e(x)$ is the electron number density. 
In our case, $n_e$ describes the Sun's and Earth's electron number profile. The complete Hamiltonian is then given by $\widetilde{H}^{(f)}=H^{(f)}+ V_W^{(f)}$, 
where the $\widetilde{}$ sign denotes the presence of a material medium. The simplest solution corresponds to the vacuum case ($VAC$), where $n_e\equiv 0$, and is given by:

\eq{\label{Pee_vacuum}
P_{ee}\left(L,E\right) &=& 1 - \sin^2 2\theta \sin^2 \left( \frac{\Delta_{G}}{4E}L\right)
}

\noindent where $L$ is the distance between the source and the detector. The resulting periodic pattern has an oscillation 
length $\lambda = 4\pi E /|\Delta_G|$, where the absolute value of $\Delta_G$ is used since it can become negative (in the $-VEP$ case). 
Actually, $L$ is fixed (a characteristic of the experiment) and we observe $P_{ee}$ as a function of $E$. If VEP is not present, $\lambda$ 
depends linearly on $E$ and the oscillation stops when $E\gg L\Delta m^2 /4\pi$. Otherwise, the dependence of $\Delta_G$ on $E$ prevents the oscillation 
from stopping as now $\lambda$ is maximum where $|\Delta_G| /E$ is minimum, at $E=E_\star$. In a particular case, $\lambda\rightarrow\infty$ 
when $E\rightarrow E_\star$, for the $-VEP$ scenario. 

\begin{figure}[ht]
		\subfloat[VAC+VEP case.]{\label{fig_vac:mais} \epsfig{file=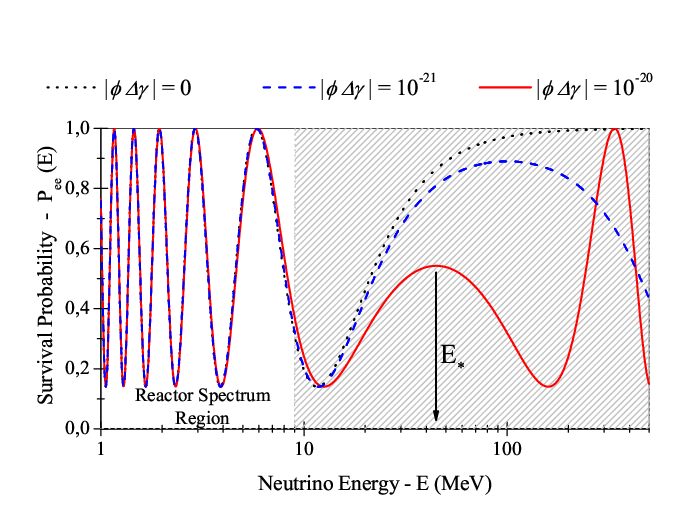,width=8.8cm}}
		\vspace{-0.5cm}
		\subfloat[VAC-VEP case.]{\label{fig_vac:menos}\epsfig{file=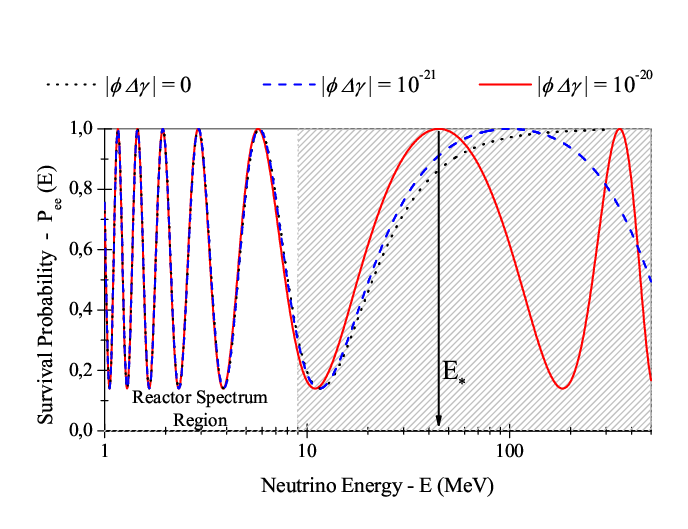,width=8.8cm}}
		\caption{\footnotesize \textsc{Survival probability $P_{ee}(E)$ for VAC$\pm$VEP}.  Fig. \ref{fig_vac:mais} corresponds to VAC+VEP and  \ref{fig_vac:menos} to VAC-VEP. The parameters corresponding to the usual neutrino mixing are taken to be $\sin^2 2\theta = 0.86$, $\Delta m^2 = 8.0\times 10^{-5}eV^2$ and $L=180km$~\citep{KL} (this last one corresponds to the average distances considered for KamLAND). Each line represents a specific value of VEP: $|\phi\ \Delta \gamma| = 0$ (no $VEP$), $10^{-21}$ ($E_\star = 140MeV$) and $10^{-20}$ ($E_\star =45MeV$).}
		\label{fig:vac}
\end{figure}

Figs. \ref{fig_vac:mais} and \ref{fig_vac:menos} show expression (\ref{Pee_vacuum}) for $+VEP$ and $-VEP$, respectively. 
The values used for $\sin^2 2\theta$ and $\Delta m^2$ are those found in the literature~\citep{PDG} for the standard Large Mixing Angle (LMA) 
solution to the solar neutrino problem\cite{PDG} and KamLAND experiment\cite{KL}, and $L$ is constant and refers to the KamLAND~\citep{KL} 
experiment average source-detector distance. In both figures, one can observe the new effect where the oscillations are restored for energies above $E_\star$. 
The presented values of $|\phi\ \Delta \gamma|$ are chosen so that any predicted new effect will not be visible within the reactor energy 
range (approximately $E\le 9MeV$). This gives us a visual ``first limit" for VEP as $|\phi\Delta\gamma|\le 10^{-20}$, if the present data reject the hypothesis.

Solutions that describe solar neutrinos must consider the Sun's matter profile, given by the Solar Standard Model (SSM)~\citep{BS05}. 
The MSW effect predicts not only conversion between flavor states but also, under certain conditions, conversion of mass states referred 
to as {\it non-adiabatic effects}~\citep{smirnov,ppal}. To better understand these effects, one has to transform $\widetilde{H}^{(f)}$ 
back to the mass basis, where it should be diagonal. The introduction of the weak potential $V_W$ assures that this transformation is 
different from $U$. It is then necessary to define the {\it effective mass basis} so that $\widetilde{\Psi}^{(m)}=\widetilde{U}^\dagger\Psi^{(s)}$ 
and $ \widetilde{H}^{(m)}=\widetilde{U}^\dagger\widetilde{H}^{(f)} \widetilde{U}$ where $\widetilde{H}^{(m)}$ has a diagonal form. 
The transformation $\widetilde{U}$ is defined as (\ref{eq:U}) with $\theta\rightarrow\widetilde{\theta}$. Requiring $\widetilde{H}^{(m)}$ to be diagonal, 
one arrives at the effective mixing in matter, 
given by

\eq{
\cos2\widetilde{\theta}(x) = \frac{\Delta_{G}\cos2\theta-A(x)}{\sqrt{\left[\Delta_{G}\cos 2\theta-A(x)\right]^2+\left(\Delta_{G}\sin2\theta\right)^2}}\ , \label{eq:cos-m}
}

\noindent where $A(x)\equiv2\sqrt{2}G_F E n_e(x)$. The \Sch equation will not retain the same form under such transformation, since $\widetilde{U}$ 
has a dependence on the position $x$. Transforming states and the Hamiltonian from the flavor to the effective mass basis results 
in $i\ \widetilde{U}^\dagger\frac{d}{dx}\widetilde{U}\ \widetilde{\Psi}^{(m)} = \widetilde{H}^{(m)}\widetilde{\Psi}^{(m)}$. 
The resulting evolution operator has additional off-diagonal terms that come from the derivative $\widetilde{U}^\dagger \frac{d}{dx}\widetilde{U}$, 
which together with the diagonal Hamiltonian $\widetilde{H}^{(m)}$ give us

\eq{\label{eq_Sch_massa_materia_matriz}
i\frac{d\ }{dx}
\widetilde{\Psi}^{(m)} =
\left(
\begin{array}{cc}
\displaystyle\widetilde{E}_1 & \displaystyle i\frac{d\widetilde{\theta}}{dx} \\
& \\
\displaystyle -i\frac{d\widetilde{\theta}}{dx} & \displaystyle\widetilde{E}_2
\end{array}
\right)
\widetilde{\Psi}^{(m)} \ ,
}

\noindent where $\widetilde{E}_1$ and $\widetilde{E}_2$ are the eigenvalues of $\widetilde{H}^{(m)}$ and $\widetilde{\theta}$ is implicitly 
given by (\ref{eq:cos-m}). The off-diagonal terms in (\ref{eq_Sch_massa_materia_matriz}) result in a non-zero probability of conversion 
between effective mass states. The intensity of these non-adiabatic effects can be measured by the relation between the diagonal and the 
off-diagonal terms of (\ref{eq_Sch_massa_materia_matriz}), such that when $\left|\frac{d\widetilde{\theta}}{dx}\right|\ll\left|\widetilde{E}_2-\widetilde{E}_1\right|$, 
the evolution operator on (\ref{eq_Sch_massa_materia_matriz}) is approximately diagonal. This condition can be summarized in the form of a {\it Adiabaticity Coefficient} 
$\Gamma(x,E)$ defined as

\eq{
\Gamma(x,E)&\equiv&\left|\frac{\frac{d\widetilde{\theta}}{dx}}{\Delta\widetilde{E}}\right|\ ,\label{eq:Gamma}
}

\noindent where 

\eq{
\Delta\widetilde{E} &=& \widetilde{E}_2-\widetilde{E}_1 \\
&=& \frac{\sqrt{\left[\Delta_{G}\cos 2\theta-A(x)\right]^2+\left(\Delta_{G}\sin2\theta\right)^2}}{2E}\ .
}

When $\Gamma(x,E)\gtrsim 1$ non adiabatic effects occur and conversion between masss eigenstates can happen.

\begin{figure}[ht]
		\subfloat[MSW+VEP case.]{\label{fig_gamma:mais}\epsfig{file=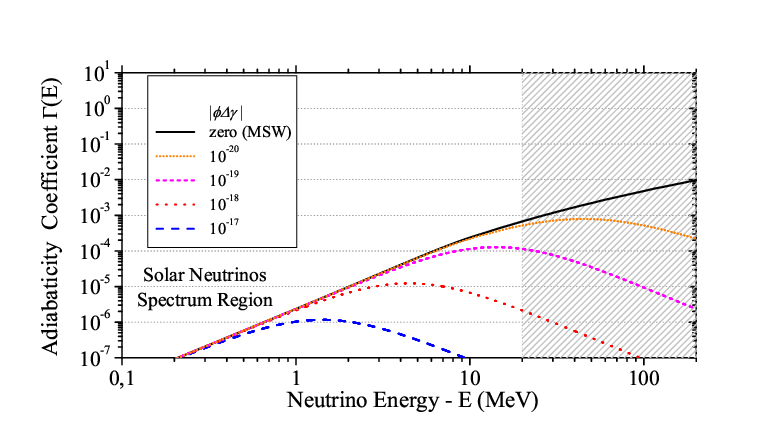,width=9.0cm}}
		\vspace{-0.1cm}
		\subfloat[MSW-VEP case.]{\label{fig_gamma:menos} \epsfig{file=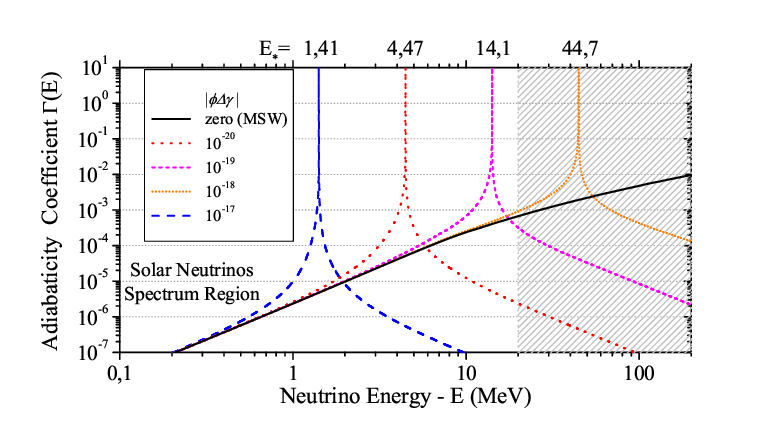,width=9.0cm}}
		\caption{\footnotesize \textsc{Adiabaticity Coefficient $\Gamma(E)$ for MSW$\pm$VEP}.  Fig. \ref{fig_gamma:mais} 
		corresponds to MSW+VEP and  \ref{fig_gamma:menos} to MSW-VEP. The parameters corresponding to the usual neutrino
		mixing are taken to be $\sin^2 2\theta = 0.86$, $\Delta m^2 = 8.0\times 10^{-5}eV^2$ and the Sun's matter profile 
		is the one from BS05(OP)~\citep{BS05}. Each line represents a specific value of VEP: 
		$|\phi\ \Delta \gamma| = 0$ (no $VEP$), $10^{-20}$, $10^{-19}$, $10^{-18}$ and $10^{-17}$. 
		For the $+VEP$ case, the system is adiabatic. For $-VEP$, non-adiabatic effects occur 
		when $E\rightarrow E_\star$ for any value of $|\phi\ \Delta \gamma|$}
\end{figure}

\noindent To better appreciate non-adiabatic effects on the neutrino spectrum, it is useful to define $\Gamma$ as a function 
of the energy only, eliminating the $x$ dependence by taking the maximum value of $\Gamma(x,E)$ for any $E$, i.e. $\Gamma(E)\equiv max\{\Gamma(x,E)\}$. 
Fig. \ref{fig_gamma:mais} shows $\Gamma(E)$ for MSW only in comparison with the MSW$+$VEP case, for some values of $|\phi\Delta\gamma|$. 
In the solar neutrino spectrum region $\Gamma(E)$ is never higher than $10^{-3}$ and so non-adiabatic effects are not expected in this case. 
On the other hand, fig. \ref{fig_gamma:menos} reveals that extremely non-adiabatic effects occur in the characteristic energies $E_\star$ 
for the MSW$-$VEP case. Such behavior happens when $\Delta_G\rightarrow 0$, causing both $\widetilde{E}_1$ and $\widetilde{E}_2$ to vanish. 
As a consequence, the off-diagonal terms of (\ref{eq_Sch_massa_materia_matriz}) become infinitely larger than the diagonal ones (as these goes to zero), 
even when $\frac{d\widetilde{\theta}}{dx}$ is naturally small, as they are expected to be in the Sun (as can be seen from $\Gamma(E)$ 
in Figs. \ref{fig_gamma:mais} and \ref{fig_gamma:menos}).

For those cases where $\Gamma(x,E)\ll 1$, equation (\ref{eq_Sch_massa_materia_matriz}) may be solved in the adiabatic approximation that leads to 
the following survival probability~\citep{smirnov}:

\eq{
P^{ad}_{ee}(x) &=& \frac{1}{2}\left[1+\cos2\widetilde{\theta}_o\cos2\widetilde{\theta}(x)\right.\nonumber \\
&+&\left.\sin2\widetilde{\theta}_o\sin2\widetilde{\theta}(x)\ \cos\alpha(x)\right]\label{Pee_msw-vep}
}

\noindent where $\cos 2\widetilde{\theta}_o$ is also given by (\ref{eq:cos-m}) where 
$\cos 2\widetilde{\theta}_o\equiv\cos 2\widetilde{\theta}(x_0)$ being $x_0$ the neutrino production point.
The factor $\cos \alpha(x)$ corresponds to the oscillating term with 

\eq{\alpha(x)=\int_0^x \Delta\widetilde{E}(x')\ dx'.}

\noindent When the matter contribution vanishes, expression (\ref{Pee_msw-vep}) corresponds to vacuum solution (\ref{Pee_vacuum}). 
Moreover, if $\Delta_G\gg10^{-10}eV^2$, $\cos \alpha(x)$ rapidly oscillates (when compared to the Sun's dimensions~\citep{bahcall,ppal}) 
and is ruled out by the average over the production point $x_0$. Without VEP this condition is satisfied as 
$\Delta m^2 = 8.0\times 10^{-5}eV^2$~\citep{PDG}, what leads to the useful simplified survival probability for solar neutrinos,

\eq{\label{Pee_ad}
P^{ad}_{ee}(x)&=& \frac{1}{2}\left[1+ \cos 2\widetilde{\theta}_o\ \cos 2\widetilde{\theta}(x)\right]\ .
}

\begin{figure}[ht]
		\subfloat[MSW+VEP case.]{\label{fig_msw:mais}\epsfig{file=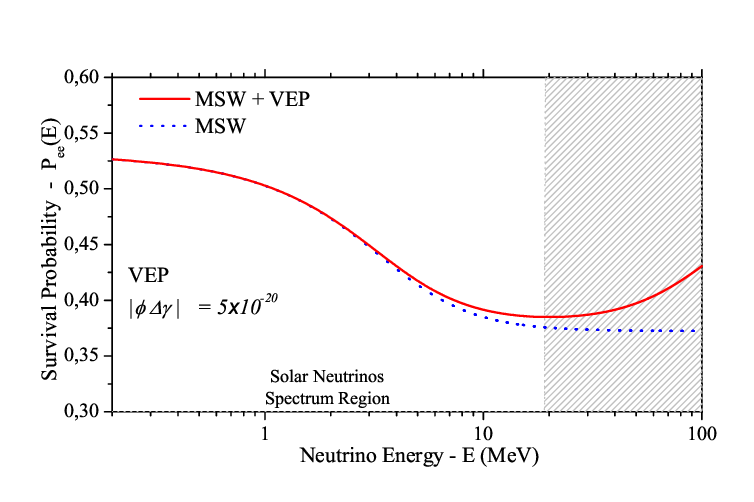,width=9.0cm}}
		\vspace{-0.1cm}
		\subfloat[MSW-VEP case.]{\label{fig_msw:menos} \epsfig{file=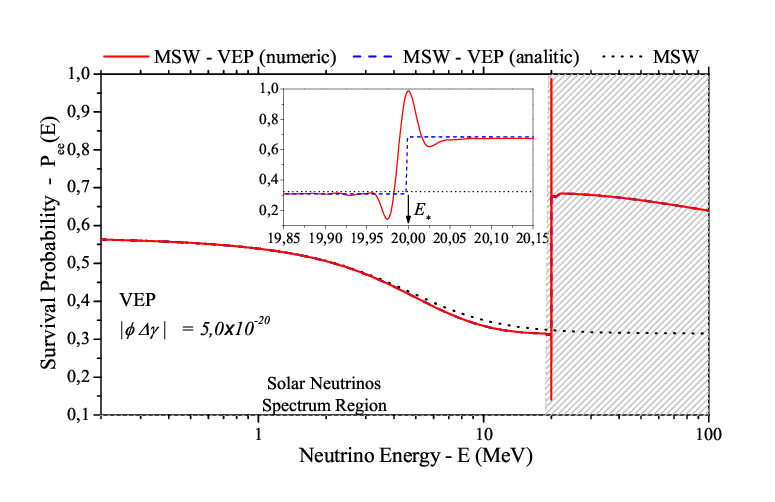,width=9.0cm}}
		\caption{\footnotesize \textsc{Survival probability $P_{ee}(E)$ for MSW$\pm$VEP}.  Fig. \ref{fig_msw:mais} corresponds to MSW+VEP and  
		\ref{fig_msw:menos} to MSW-VEP. The parameters corresponding to the usual neutrino mixing are taken to be 
		$\sin^2 2\theta = 0,86$, $\Delta m^2 = 8,0\times 10^{-5}eV^2$. On fig \ref{fig_msw:mais}, expression (\ref{Pee_ad}) 
		compares the case without VEP and for $|\phi\Delta\gamma| = 5\times 10^{-20}$. Fig. \ref{fig_msw:mais} also compares a numerical 
		solution for (\ref{eq_Sch_massa_materia_matriz}) that shows new non-adiabatic effects in the proximities of $E_\star$ (in detail).}
		\label{fig:msw}
\end{figure}

From the studies of the adiabaticity coefficient, the above expression is expected to hold for the $MSW+VEP$ case and almost everywhere for $MSW-VEP$, 
except in the neighborhood of $E_\star$. Fig. \ref{fig_msw:mais} shows a comparison between expression (\ref{Pee_ad}) with no VEP and 
with $|\phi\Delta\gamma| = 5\times 10^{-20}$, for the $MSW+VEP$ case. This value of VEP was chosen so that $E_\star$ is just above the solar neutrino spectrum
($E_\star = 20MeV$). The survival probability with VEP is always greater than the one for MSW only, but this difference only becomes appreciable 
for energies above $E_\star$. Fig. \ref{fig_msw:menos} shows the same comparison for MSW$-$VEP. As expected, non-adiabatic effects occurs near $E_\star$. 
Fig. \ref{fig_msw:menos} also shows a numerical solution of the equation (\ref{eq_Sch_massa_materia_matriz}), in which the non-adiabatic behavior can be seen in details. 
These effects are confined inside a narrow region around $E_\star$ and they are not observable with the present data statistics. The adiabatic approximation 
(\ref{Pee_ad}) describes well the survival probability by any practical means, in both $\pm$VEP cases.

\begin{figure}[ht]
		\epsfig{file=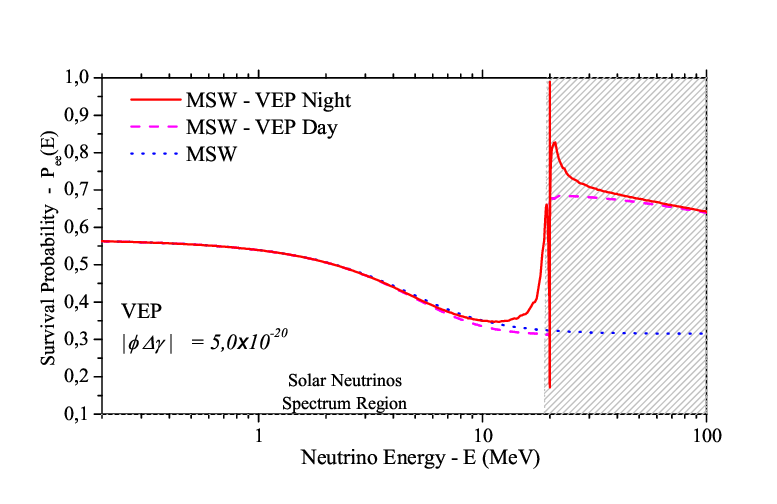,width=9.0cm}
		\caption{\footnotesize \textsc{Survival probability $P_{ee}(E)$ for neutrinos arriving at night, with MSW$-$VEP}. 
		A comparison between day and night probabilities shows an excess for the night time. The parameters used in this plot are the same as in 
		Fig. \ref{fig_msw:menos}. }\label{fig_msw:menos-noite}
\end{figure}

\begin{figure}[ht]
		\epsfig{file=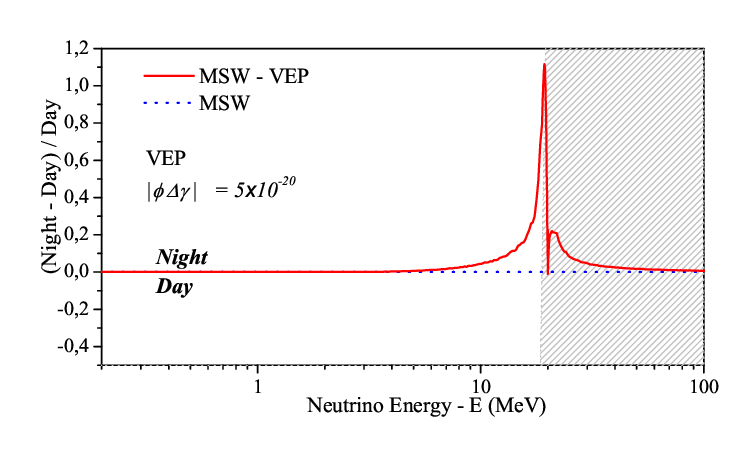,width=9.0cm}
		\caption{\footnotesize \textsc{Day-Night asymmetry for MSW$-$VEP}. 
		The $-VEP$ hypothesis predicts new regeneration effects due to Earth's matter generating an excess of solar neutrinos arriving at night time, 
		for energies close to $E_\star$.
		}\label{fig_msw:dia-noite}
\end{figure}

At night time, solar neutrinos cross several Earth layers. Again, the presence of matter alters the survival probability in a way that night neutrinos 
have more chance to survive than those arriving at day. This effect is called {\it regeneration} and is not observed on the solar neutrino data~\citep{holanda1}. 
On the same way that new non-adiabatic effects were predicted for day neutrinos in the $MSW-VEP$ case, new regeneration signal may also be expected. 
To account for these possibilities a numerical solution of (\ref{eq_Sch_massa_materia_matriz}), for the Earth's matter profile, is used. Fig. \ref{fig_msw:menos-noite}
is the equivalent to \ref{fig_msw:menos} for night neutrinos. In the neighborhood of $E_\star$, the solar non-adiabatic effects are intensified by Earth's matter. 
Fig. \ref{fig_msw:dia-noite} shows a measure of the asymmetry between night and day for the $-VEP$ case. As it can be seen, an excess is expected in a 
region wider than the one where the solar non-adiabatic effect takes place. The absence of regenerations signs on the solar neutrinos data imposes a 
stronger limit on $E_\star$ for the $-VEP$ case than for $+VEP$. This has direct consequences on the limits for $|\phi\Delta\gamma|$.

\section{Data Analysis}

For solar neutrinos, we consider data from Homestake ~\citep{HOMESTAKE},
Sage~\citep{SAGE}, Gallex/GNO~\citep{GNO}, SuperKamikande(SK)~\citep{SK},
SNO (I~\citep{SNOI}, II~\citep{SNOII} and III~\citep{SNOIII}) and Borexino~\citep{BOREXINO, BOREXINO2010} 
experiments. As for reactor anti-neutrinos, KamLAND data is considered~\citep{KL}. A $\chi^2$
analysis is done, where we define

\eq{ \chi^2=\chi^2_{sun}+\chi^2_{KL}.  }

\noindent The solar neutrinos contributions $\chi^2_{sun}$ is given by
~\citep{holanda1}:

\eq{\label{X2S} \chi^2_{sun}= \sum_{i,j=1}^{119}
  \left[R^{th}_i-R^{ex}_i\right]\left[S^2\right]^{-1}_{ij}\left[R^{th}_j-R^{ex}_j\right]\ ,
}

\noindent where $R^{th}_i$ and $R^{ex}_i$ are the theoretical and
experimental rates respectively and $S^2$ takes into account all
the correlation between uncertainties~\citep{Fogli, SNOII}. Reactor
neutrinos contribute through a Poisson statistics~\citep{holandaKL}:

\eq{\label{X2KL}
  \chi^2_{KL}=\sum_{i=1}^{24}2\left[N_i^{th}-N_i^{ex}+N_i^{ex}\ln\left(\frac{N_i^{ex}}{N_i^{th}}\right)\right]
}

\noindent where $N_i^{th}$ and $N_i^{ex}$ are the theoretical and experimental counting. 
For our purposes in this Latter, which are seeking for limits on VEP parameters, 
this definition of $\chi^2_{KL}$ is enough, even if it is not taking in account the systematic and correlated errors related to the KamLAND statistics.

\begin{figure}[ht]

		\vspace{0.6cm}
		\subfloat[MSW+VEP case.]{\label{fig:vepmais.tgdm}\epsfig{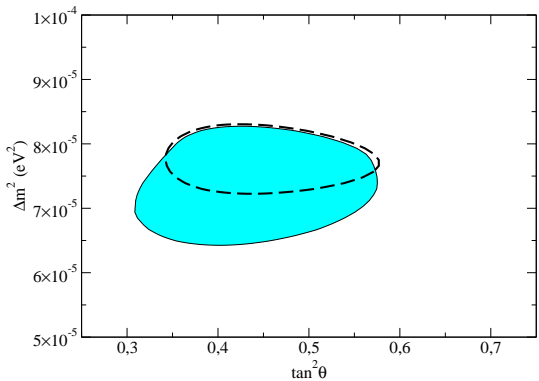}}

		\vspace{0.5cm}
		\subfloat[MSW-VEP case.]{\label{fig:vepmenos.tgdm} \epsfig{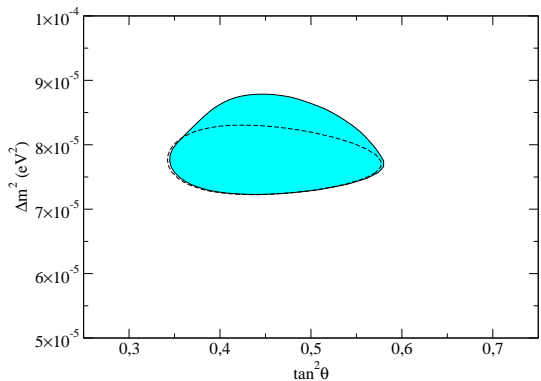}}
		\caption{\footnotesize \textsc{Comparison between pure $MSW$(LMA) solution and $MSW\pm VEP$}. $\chi^2$ 
		maps in the $\Delta m^2\times \tan^2\theta$ plane, with $|\phi\Delta\gamma|$ minimized on every point. 
		On both figures, the shaded area shows the $3\sigma$ region for $MSW\pm VEP$, while the dashed line shows $MSW$ only.}
		\label{fig:chi2}
\end{figure}

\begin{figure}[ht]
		\subfloat[MSW+VEP: $|\phi\Delta\gamma|\times \tan^2\theta$]{\label{fig:vepmais.tgvep} \epsfig{file=chi2.vepplus.tgvep.eps,width=7.0cm}}

		\vspace{0.5cm}
		\subfloat[MSW+VEP: $\Delta m^2\times|\phi\Delta\gamma|$]{\label{fig:vepmais.vepdm} \epsfig{file=chi2.vepplus.vepdm.eps,width=7.0cm}}
		\caption{\footnotesize \textsc{Limits for $|\phi\Delta\gamma|$ in the $+VEP$ case.} 
		Both figures show $\chi^2$ maps, in the $|\phi\Delta\gamma|\times \tan^2\theta$ (Fig.\ref{fig:vepmais.tgvep}) 
		and $\Delta m^2\times|\phi\Delta\gamma|$ (Fig.\ref{fig:vepmais.vepdm}) planes. 
		The dotted and dashed lines indicate the limits coming from KamLAND and Solar neutrinos, 
		respectively. All curves correspond to $3\sigma$.}
		\label{fig:chi2.vepmais}
\end{figure}

\begin{figure}[ht]
		\subfloat[MSW-VEP: $|\phi\Delta\gamma|\times \tan^2\theta$]{\label{fig:vepmenos.tgvep} \epsfig{file=chi2.vepminus.tgvep.eps,width=7.0cm}}

		\vspace{0.5cm}
		\subfloat[MSW-VEP: $\Delta m^2\times|\phi\Delta\gamma|$]{\label{fig:vepmenos.vepdm} \epsfig{file=chi2.vepminus.vepdm.eps,width=7.0cm}}
		\caption{\footnotesize \textsc{Limits for $|\phi\Delta\gamma|$ in the $-VEP$ case.}Both figures 
		show $\chi^2$ maps, in the $|\phi\Delta\gamma|\times \tan^2\theta$ (Fig.\ref{fig:vepmenos.tgvep}) 
		and $\Delta m^2\times|\phi\Delta\gamma|$ (Fig.\ref{fig:vepmenos.vepdm}) planes. 
		The dotted and dashed lines indicate the limits coming from KamLAND and Solar neutrinos, respectively. 
		All curves correspond to $3\sigma$.}
		\label{fig:chi2.vepmenos}
\end{figure}

The results of our statistical analysis are shown in Figs. \ref{fig:chi2}, \ref{fig:chi2.vepmais} and \ref{fig:chi2.vepmenos}. 
In Fig. \ref{fig:chi2}, a comparison between pure MSW(LMA) solution (region inside dashed line) and the solution including both 
MSW and VEP effects (shaded area) are shown. From Fig. \ref{fig:vepmais.tgdm} we observe that the inclusion of $+VEP$ effects makes 
the $3\sigma$ compatibility region move towards lower values of the $\Delta m^2$ parameter, while the $\tan^2\theta$ remains almost unchanged. 
This is a consequence of the fact that the inclusion of a positive number proportional to $|\phi\Delta\gamma|$ to $\Delta m^2$ allows lower 
values of $\Delta m^2$, as can be appreciated through eq. \ref{Delta_Gpm}. The opposite situation happens when $-VEP$ effects are considered, 
and Fig. \ref{fig:vepmenos.tgdm} reflects it. Figs. \ref{fig:vepmais.tgdm} and \ref{fig:vepmenos.tgdm} also show the 
limits coming from KamLAND and Solar data alone, in each case.

For both $+VEP$ and $-VEP$ cases, the analysis shows that the standard global solution for solar and reactor neutrinos (MSW/LMA) 
is recovered when $|\phi\Delta\gamma|\le 10^{-21}$. On the other hand, $VEP$ effects start to be significant when $|\phi\Delta\gamma|$ is just above $10^{-20}$. 
This range is shown on figures \ref{fig:chi2.vepmais} and it is consistent with figures \ref{fig:vac} and \ref{fig:msw}. 
The superior limits obtained for each case are: $|\phi\Delta\gamma|< 9.0\times 10^{-20}$ ($3\sigma$) for $-VEP$ and $|\phi\Delta\gamma|\le 2.0\times 10^{-19}$ ($3\sigma$) 
for $+VEP$. Although both limits are very similar, the superior limit for presence of VEP, regardless of the sign case, is the less restrictive of these values:

\begin{equation}\label{upperbound}
|\phi\Delta\gamma|< 2.0\times 10^{-19}(3\sigma),
\end{equation}

\noindent since the $+$ and $-VEP$ cases are obviously mutually excluding. 

In the specific case of atmospheric or large base-line $\nu_\mu$  neutrinos, a different mass difference scale is involved in such a 
way that one only needs to consider for $\nu_\mu \rightarrow \nu_\tau$  oscillations. So, one expects that at large enough energies 
any effect of VEP should be dependent only on $\Delta\gamma_{23}$. 

A limit on this VEP scale has already been obtained in \citep{Concha}:  $|\phi\Delta\gamma_{23}| = 6.3\times10^{-25}$. 
This limit comes from the $\nu_\mu \rightarrow \nu_\tau$ channel, with energies of the order of or greater than the GeV scale. 
So, one would not expect that any possible VEP effect below this limit could influence our results, even in a three flavor scenario. 
This can be appreciated by looking at the characteristic energy scale, which for this case is $E_{\star 23}>30GeV$. Even if one considers this  
Atmospheric/Accelerator channel on the Solar/Reactor analysis, any possible VEP effect coming from this sector is lower bounded in energy to a scale 
well over the Solar/Reactor one ($E<15MeV$).

On the opposite direction, one could ask also if the VEP parameter $|\phi\Delta\gamma_{12}|$ can be constrained by data from Atmospheric 
and/or Accelerator observations. A very naive estimate can be done even in the three neutrino analyses. The VEP effects with neutrino 
energies much higher than solar and reactor ones would lead to a phenomenological situation corresponding to   
$\Delta m_{23}<<\Delta m_{12}$ and $\Delta m_{23} <<\Delta m_{13}$. This would imply that when we assume $\phi\Delta\gamma_{12}$ 
in the limit we have found for this VEP parameter shown in the eq.(\ref{upperbound}), a normalization factor over the usual two oscillation 
scenario is found: $P_{\mu\mu}\approx 0.74 P^{no VEP}_{\mu\mu}$. In a first approximation, atmospheric observations can be ignored in the 
present Letter because the flux of atmospheric neutrinos are obtained within uncertainties of order of 25\%~\cite{Honda2007} which can be 
absorbed by the normalization factor. Therefore,  we do not expect to find strong consequences on 
the constraints of VEP parameters\footnote{Note also that after the conclusion of this manuscript, MINOS established limits on the electronic neutrino 
appearance\cite{MINOS2011} which could impose some limit on $\phi\Delta\gamma_{12}$. Nevertheless, 
in the same way as discussed for atmospheric neutrinos, the $\nu_e$-survival probability can be 
naively calculated to be around $0.2$, which is the same order of magnitude of the observed limit of appearance of $\nu_e$ in this experiment. 
This means that no significant limit on the VEP parameter we are interested in will appear. A deeper analysis of this situation will be done in the future\cite{valdiviesso}.}. 
A detailed analysis of this case is in preparation\cite{valdiviesso}. 


In order to compare our limits on VEP parameters with the ones coming from other macroscopic experiments~\citep{Su,Smith}, 
one has to consider an estimative of the gravitational potential $\phi$. It seems that, among several possible sources, 
the {\em Great Attractor} offers the largest contribution~\citep{Halprin2}, with its  best estimative given by $\phi=3\times 10^{-5}$. 
So the upper bound of $|\phi\Delta\gamma|$ given in (\ref{upperbound}) corresponds to a maximum value of the order $|\Delta\gamma|< 10^{-14}$.

\section{Conclusion}\label{conclusao}

The results of our analysis imposes a new limit for the Violation of the Equivalence Principle. 
The model offers two theoretical possibilities: one in which greater mass represents greater gravitational 
coupling (here called $VAC/MSW+VEP$) and an inverse situation, where great\-er mass implies a smaller coupling with the gravitational field ($VAC/MSW-VEP$). 
With latest statistics presented by the KamLAND collaboration and all solar neutrino data, we obtained a new limit for the VEP of $1$ part in $10^{14}$ 
in neutrino oscillation channels involving $\nu_e$ and $\bar{\nu}_e$ disappearance. This limit should be carefully compared with different limits previously obtained. 
Macroscopic experiments imposed limits of $1$ part in $10^{12}$ for VEP~\citep{Su} and neutrino experiments based on different oscillation channels, 
specifically $\nu_\mu\rightarrow\nu_\tau$, imposed limits of $1$ part in $10^{20}$~\cite{Concha}.

A final comment is in order. The VEP hypothesis presented here is just one possible option. Any model that presents a mixing scenario,
with a Hamiltonian like (\ref{H_massa2}) and with $\Delta E$ given by an expression with the same momentum dependency as the one seen in (\ref{DeltaE}),
would be limited by the same values just obtained. The combination of Violation of Lorentz Invariance (VLI) models~\citep{Arias} with mass-flavor mixing 
presents the same phenomenological behavior as shown in (\ref{H_massa2}) and (\ref{DeltaE}), needing only a parameter reinterpretation: $\Delta c=2|\phi|\Delta\gamma$, 
where $\Delta c=c_2-c_1\neq 0$ implies VLI between neutrino flavors (as defined on section 2 of~\citep{Arias}), being $c_1$ and $c_2$ the limiting speeds for 
two different neutrino mass eigenstates. So this work also imposes a limit in this parameter: $|\Delta c|\le 4\times 10^{-19}$ (for the solar sector).

We would like to thank FAPESP, CNPq and CAPES for several financial supports.

\bibliography{bibliografia}

\end{document}